\documentstyle[12pt]{article}
\renewcommand {\c}  {\'{c}}
\newcommand {\cc} {\v{c}}
\newcommand   {\s}  {\v{s}}
\begin{document}
\begin{titlepage}
\begin{flushright}
{\bf RBI-TH-3/94}\\
{\bf PMF-ZTF-3/94}\\
{\bf February 1994}
\end{flushright}
\pagestyle{empty}
\vspace* {13mm}
\baselineskip=24pt
\begin{center}
%
  {\bf UNIFIED VIEW OF  DEFORMED SINGLE - MODE  OSCILLATOR ALGEBRAS  }
  \\[10mm]
S.Meljanac$^1$,  M.Milekovi\c$^{2,+}$ and S.Pallua$^{2}$\\[7mm]
{\it $^1$ Rudjer Bo\s kovi\c \ Institute, Bijeni\cc ka c.54, 41001 Zagreb,
Croatia\\[5mm]
 $^2$ Prirodoslovno-Matemati\cc ki Fakultet, Zavod za teorijsku fiziku,
\\Bijeni\cc ka c.32, 41000 Zagreb, Croatia\\[5mm]
$^+$ e-mail: marijan@phy.hr}\\
\vskip 1cm
{\bf Abstract}
\end{center}
\vskip 0.2cm
A general framework for the deformation of the single-mode oscillators is
presented and all deformed single-mode oscillators are unified.
The extensions of the Aric-Coon, genon, the para-Bose and the para-Fermi
oscillators are proposed. The generalized harmonic oscillator considered by
Brzezinski et al.
is rederived in a simple way.Some remarks on  deformation of $SU(1,1)$ and
supersymmetry
are made.\\
\begin{center}
{\it ( to appear in Phys.Lett.B )}
\end{center}
\end{titlepage}
%
%
%
\newpage
\setcounter{page}{1}
\pagestyle{plain}
\def\leer{\vspace{5mm}}
\baselineskip=24pt
\setcounter{equation}{0}%
Recently, different partial results for the deformed single-mode
oscillator have been obtained\cite{ac}-\cite{jag}. Our motivation is to
present a unified view of all known deformations and to discuss some
extensions. In particular, we consider extensions of the
Arik-Coon\cite{ac}, para-Bose, para-Fermi oscillators\cite{ok} and
their deformations\cite{okk}\cite{fv}\cite{bon}. We show that even cases which
were previously discarded as unphysical,
owing to the negative norms\cite{ac}\cite{g}, can be acceptable deformations
of single-mode oscillators. The "genon" oscillators\cite{cd} are also included.
Finally, we discuss the most generally deformed $SU(1,1)$ algebra and
comment on hidden supersymmetry. The simplicity of our approach is
demonstrated on all known results \cite{ac}-\cite{jag}. \\
Let us consider a pair of operators $\bar{a}$, $a$ (not necessarily hermitian
conjugate to each other) with the number operator N. The most general
commutation relation linear in the $\bar{a} a$ and $a \bar{a}$ operators is
\begin{equation}
  a \bar{a} - F(N) \bar{a} a = G(N),
\end{equation}
where $F(N)$ and $G(N)$ are arbitrary complex functions. The number
operator satisfies
\begin{equation}
\begin{array}{c}
[N,a] = -a,\\[4mm]
[N,\bar{a}] = \bar{a},\\[4mm]
[N, \bar{a} a] = [N,a \bar{a}] = 0 .
\end{array}
\end{equation}
Hence, we can write
\begin{equation}
\begin{array}{c}
\bar{a} a = \varphi (N),\\[4mm]
a \bar{a} =  \varphi (N+1),
\end{array}
\end{equation}
where $\varphi (N)$ is, in general, a complex function satisfying
the recurrence relation
\begin{equation}
\varphi (N+1) - F(N) \varphi (N) = G(N).
\end{equation}
If $\varphi (N)$ is the bijective mapping, then
\begin{equation}
N = \varphi^{-1}(\bar{a} a) = \varphi^{-1}(a \bar{a}) - 1.
\end{equation}

Let us denote  the hermitian conjugate of the operator
$a$ by  $a^{\dagger}$ . Then it follows that
\begin{equation}
\begin{array}{c}
[N,a^{\dagger}] = a^{\dagger},\\[4mm]
\bar{a}  = c(N) a^{\dagger},
\end{array}
\end{equation}
where $c(N)$ is a complex function of $N$. It is convenient to choose
$c(N)$ to be a "phase" operator, $|c(N)| = 1$. Then we have
\begin{equation}
\begin{array}{c}
a^{\dagger} a = |\varphi (N)|,\\[4mm]
a a^{\dagger} = |\varphi (N+1)|,\\[4mm]
a a^{\dagger} - a^{\dagger} a  =  |\varphi (N+1)| - |\varphi (N)| =
G_{1}(N),\\[4mm]
c(N) = e^{i \arg \varphi (N)} = \frac{\varphi (N)}{|\varphi (N)| }.
\end{array}
\end{equation}
If $\varphi (N) > 0$, then $\arg \varphi (N) = 0$ and $c(N) = 1$.\\
Let us further assume that $|0>$ is a vacuum:
 \begin{equation}
\begin{array}{c}
a |0> = 0,\\[4mm]
N |0> = 0 , \quad    \varphi (0) = 0,\\[4mm]
<0|0> = 1 .
\end{array}
\end{equation}
One can always normalize the operators $a$ and $\bar{a}$ such that
$|\varphi (1)| = 1$; then\\ $|G(0)| = 1$. The function $\varphi (N)$ is
determined by the recurrence relation (4) and is given by
\begin{equation}
\varphi (n) = [F(n-1)]! \sum_{j=0}^{n-1} \frac{G(j)}{[F(j)]!},
\end{equation}
where
\begin{equation}
\begin{array}{c}
[F(j)]! = F(j)F(j-1)...F(1),\\[4mm]
[F(0)]! = 1 .
\end{array}
\end{equation}
The excited states with unit norms are
$$
|n> = \frac{(a^{\dagger})^{n}}{\sqrt {[|\varphi (n)|]!}} |0> =
\frac{(c^{-\frac{1}{2}} \bar{a})^{n}}{\sqrt {[\varphi (n)]!}} |0>,\\[4mm]
$$
\begin{equation}
\begin{array}{c}
<n|m> = \delta_{mn},   \quad   n,m = 0,1,2....,\\[4mm]
<n-1|a|n> = <n|a^{\dagger}|n-1> = \sqrt {|\varphi (n)|} .
\end{array}
\end{equation}
If $\varphi (n) \neq 0$ for $\forall n$ $ \in$ ${\bf N}$, then there is an
infinite
set ("tower") of states. However, if  $\varphi (n_{0}) = 0$ for some $n_{0}$,
 then the state $(a^{\dagger})^{n_{0}} |0>$ has zero norm and, consistently,
we can put $|n_{0}> \equiv  0$. The corresponding representation is
finite-dimensional and the representation matrices are of the
$n_{0} \times n_{0}$ type ((11)).\\
The operators $a$ and $a^{\dagger}$ can be related to the Bose operators
$b$ and $b^{\dagger}$ by mapping
\begin{equation}
a = b \sqrt{\frac{|\varphi (N)|}{N}},  \quad    a^{\dagger} =
\sqrt{\frac{|\varphi (N)|}{N}} b^{\dagger},
\end{equation}
where
\begin{equation}
b b^{\dagger} -  b^{\dagger} b =1,  \quad  N = b^{\dagger} b.
\end{equation}
Note that this transformation preserves the number operator, i.e.
$N^{(a)}=N^{(b)}$.\\
If $\varphi (n) \neq 0$ for $\forall n \in {\bf N}$, the Fock space of
the deformed algebra is identical  to the Fock space of the Bose
oscillator. However, if $\varphi (n) = 0$ for some $n_{0} \in {\bf N}$, the
Fock space of the Bose oscillator reduces to finite dimensional subspace.
 Then $(a^{\dagger})^{n} |0> = 0$ for some $n \geq n_{0}$.\\
Note that  we
can define a new vacuum for every $n_{0} \in {\bf N}$, for which $\varphi
(n_{0}) = 0$:
$$
|n_{0}> = \frac{(b^{\dagger})^{n_{0}})}{\sqrt {(n_{0})!}} |0>,
$$
\begin{eqnarray}
a |0) \equiv a |n_{0}> =b \sqrt{\frac{|\varphi
(N)|}{N}}\frac{(b^{\dagger})^{n_{0}})}{\sqrt {(n_{0})!}} |0> = 0, \nonumber \\
a^{\dagger} |0) \equiv a^{\dagger} |n_{0}> = \sqrt{|\varphi
(n_{0}+1)|}|n_{0}+1> \neq 0 \label{}.
\end{eqnarray}
The vacuum $|0)$ is identical to the $n_{0}$-particle state and
$|1)$ corresponds to the new one-particle state. Hence, we can define
as many vacua as there are solutions of the equation $\varphi (n_{0})
 = 0$, $n_{0} \in {\bf N}$.The Fock space of the Bose oscillator is split into
subspaces corresponding to different vacua. The new number operator
$\aleph $ is defined as $\aleph |0) = \aleph |n_{0}> = 0$,
$\aleph = N - n_{0}$.\\
The new excited states are
\begin{equation}
|n)= |n+n_{0}> = \frac{(a^{\dagger})^{n}}{\sqrt{|\varphi (n+n_{0})|....|\varphi
(n_{0}+1)|}} |0),
\end{equation}
$$
\aleph = N - n_{0} .
$$
Different vacua are not neccessarily degenerate. Their relative positions
depend on the hamiltonian
expressed in terms of $a$ and $a^{\dagger}$. (For example, if $H=a^{\dagger}a$,
then  all the vacua are degenerate).
Note that $(a^{\dagger})^{n}$ $\neq 0$  for $\forall n$ $\in$ ${\bf N}$ in the
case of different vacua  .\\
The function $|\varphi (n)|$ uniquely determines the type of deformed
oscillator algebra and vice versa. If $\varphi _{1}$ $\neq$ $\varphi _{2}$ but
 $|\varphi _{1}| = |\varphi _{2}|$, the corresponding algebras are isomorphic.
There is a family of  functions $(F,G)$ leading to the same algebra, with
identical
 functions $\varphi (N)$. Therefore we can fix the "gauge", for example:
\begin{equation}
\begin{array}{c}
(a)  \quad       F(N) = 1 ,    \quad    a \bar{a} - \bar{a} a =
G_{1}(N),\\[4mm]
(b)   \quad     G(N) = 1,     \quad   a \bar{a} - F_{1}(N)\bar{a} a = 1,
\\[4mm]
(c)    \quad     F(N) = q,      \quad   a \bar{a} - q \bar{a} a = G_{q}(N) .
\end{array}
\end{equation}
The connection between cases (a), (b) and (c) is
\begin{equation}
G_{1}(N) = G_{q}(N) + (q-1) \varphi (N) = 1 + [F_{1}(N)-1] \varphi (N).
\end{equation}

Let us examine case (c). We assume that  $G_{q}(N)$ can be expanded
in powers of $N$ around any $n \in {\bf N}$. Then
\begin{eqnarray}
a \bar{a} - q \bar{a} a = G_{q}(N) = \sum_{k=0}^{\infty} c_{k} N^{k},  \quad
c_{0} = 1,\nonumber \\
\varphi (n) = q^{n-1}  \sum_{j=0}^{n-1} \frac {G_{q}(j)}{q^{j}} = q^{n-1}
\sum_{k=0}^{\infty} c_{k} S_{k}(n,q), \label{}
\end{eqnarray}
where
\begin{equation}
S_{k}(n,q) = \sum_{j=0}^{n-1} \frac {j^{k}}{q^{j}} \equiv q^{1-n} \{
\sum_{j=0}^{n-1}j^{k} \}_{(k,q)} .
\end{equation}
The recurrence relations for $S_{k}(n,q) $ are
\begin{eqnarray}
(-q )^{k} \frac {d^k}{dq^k} S_{0}(n,q) = \sum_{l=1}^{k} e_{l}^{(k)} S_{l}(n,q),
\nonumber \\
S_{k}(n,q) = (-q\frac {d}{dq}) S_{k-1}(n,q), \label{}
\end{eqnarray}
where
\begin{equation}
e_{l}^{(k)} = \frac {1}{l!} [\frac {d^l}{dx^l} x(x+1)...(x+l-1)]_{x=0}.\\[4mm]
\end{equation}
{}From  eq. (20) it follows that
\begin{equation}
S_{k}(n,q) =  (-q  \frac {d}{dq})^{k} S_{0}(n,q) = \sum_{l=1}^{k} e_{l}^{(k)}
S_{l}(n,q)
\end{equation}
and from eq. (18)
\begin{equation}
\varphi (n) = q^{n-1} G_{q}(-q  \frac {d}{dq}) S_{0}(n,q),
\end{equation}
where
\begin{equation}
 S_{0}(n,q) = q^{1-n} \frac{q^n - 1}{q - 1} = [n]_{q^{-1}} .
\end{equation}
Specially,  for  case (a) we obtain
\begin{equation}
\varphi (n) = \sum_{j=0}^{n-1} G_{1}(j) = [G_{1}(-q \frac{d}{dq})
S_{0}(n,q)]_{q=1}.
\end{equation}
Using these formulas, all
kinds of deformed oscillator can be unified and the notion of deformation
can also be  extended to  cases in which  states with negative norms
appeared.\\
In the following we discuss some examples.

(I) Extension of the Arik-Coon algebra:
\begin{equation}
\begin{array}{c}
aa^{\dagger} - q a^{\dagger}a = 1,\\[4mm]
\varphi (n) = \frac{q^n - 1}{q - 1},
\end{array}
\end{equation}
The algebra is defined for $q \geq -1$, since for all other $q$,  $a^{\dagger}a
= \varphi (n)$
is negative. However, we propose the following deformed algebra instead of eq.
(26),
$$
a \bar{a} - q \bar{a} a = 1 ,   \quad  q \in {\bf C},\\[4mm]
$$
$$
\bar{a} a  = \frac {q^n - 1}{q - 1} ,\\[4mm]
$$
\begin{equation}
a^{\dagger} a  = |\frac {q^n - 1}{q - 1}| = \frac {|q|^{2n} - 2 |q|^n \cos
n\varphi + 1}
{|q|^2 - 2 |q| \cos \varphi + 1}.
\end{equation}
This is a generalization of the genon oscillator \cite{cd} with $q = e^{i\frac
{2\pi}{M}}$ and
 $M \in {\bf N}$.\\
If $q < -1$, we find that
$$
a^{\dagger} a  = \frac {|q|^n + (-)^{n-1}}{|q| + 1} > 0,  \quad   \forall n \in
{\bf N},\\
$$
\begin{equation}
aa^{\dagger} - |q| a^{\dagger}a = (-1)^N .
\end{equation}
When $q=-1$, it corresponds to the Fermi oscillator. Similarly, the algebra \\
$ aa^{\dagger} + a^{\dagger}a=(-1)^N $ corresponds to the Bose oscillator.

Let us in the same way extend single para-Bose and para-Fermi oscillators
\cite{ok}
as well as their deformations\cite{okk}\cite{cs}\cite{fv}\cite{bon}.

(II) Extension of the single para-Bose oscillator
\begin{equation}
\begin{array}{c}
a \bar{a} + \bar{a} a = 1 + c_{1} N ,  \quad  c_{1} \in {\bf C},\\[4mm]
\varphi (n) = (-)^{n-1} \{ (1- c_{1} q\frac {d}{dq})\frac {q^{-n} - 1}{q^{-1} -
1}\}_{q=-1} \\[4mm]
=   (-)^{n-1} \{ S_{0}(n,-1) + c_{1} S_{1}(n,-1)\} = \frac{1}{2} c_{1}n +
[n]_{-1} (1 - \frac{1}{2} c_{1}) .
 \end{array}
\end{equation}
For $c_{1} = 0$, eq.(29) corresponds to the  Fermi oscillator. When $c_{1} =
\frac{2}{p}$, $p \in {\bf N}$
eq.(29) corresponds to the para-Bose oscillator of the $p^{th}$ order
\cite{cs}\cite{ok}\cite{bon}, and the representation is
infinite-dimensional. When  $c_{1} = -\frac{1}{p}$, $p \in {\bf N}$,  the
representations
are finite dimensional but different from the para-Fermi  oscillator.(If we
admit a different vacuum , $|0) = |p>$,
there is an infinite tower of  states on it).\\
Different deformations of a single para-Bose oscillator are proposed in
\cite{okk}\cite{fv}.
For example:
\begin{equation}
\varphi^{(a,b)} (n) = [\frac{1}{2} c_{1}n + [n]_{-1} (1 - \frac{1}{2}
c_{1})]^{(a,b)}_{q},
\end{equation}
where
$$
[x]^{(a)}_{q} = \frac {q^x - 1}{q - 1},
$$
\begin{equation}
[x]^{(b)}_{q} = \frac {q^x - q^{-x}}{q - q^{-1}} = q^{1-x} [x]^{(a)}_{q^2} .
 \end{equation}
Note that the single para-Bose algebra (29) can also be  written as
\begin{equation}
a \bar{a} - \bar{a} a =\frac{1}{2} c_{1} + (1 - \frac{1}{2} c_{1}) (-)^N .
\end{equation}
This is equivalent to the so-called "modification" \cite{bp}\cite{bmf}
\begin{equation}
aa^{\dagger} -  a^{\dagger}a = 1 + 2\nu (-)^N .
\end{equation}
If we change the operators $a$ and $a^{\dagger}$ in eq.(32) into  $a
\rightarrow a\sqrt{1+\nu}$
 and put  $\frac{1}{1+2\nu} = \frac{1}{2} c_{1}$, we obtain eq.(33).\\
Futhermore, the following deformation is proposed in ref.\cite{bmf}  :
\begin{equation}
aa^{\dagger} - q a^{\dagger}a = q^{-N} (1 + 2\nu (-)^N) .
\end{equation}
Using our equation (18) we obtain
\begin{equation}
\begin{array}{c}
\varphi (n) = q^{n-1} (S_{0}(n,q^2) + 2\nu S_{0}(n,-q^2)\\[4mm]
= q^{1-n} \{ [n]^{(a)}_{q^2} + 2\nu (-)^{n-1} [n]^{(a)}_{-q^{2}}\}\\[4mm]
= [n]^{(b)}_{q} + 2\nu (i)^{1-n} [n]^{(b)}_{iq} .
 \end{array}
\end{equation}

(III) Extension of the single para-Fermi oscillator:
\begin{equation}
\begin{array}{c}
a \bar{a} - \bar{a} a = 1 + c_{1} N ,  \quad  c_{1} \in {\bf C}\\[4mm]
\varphi (n) =  n (1 + \frac{1}{2} c_{1}(n-1)) .
 \end{array}
\end{equation}
For $c_{1}=0$, eq.(36) corresponds to the Bose oscillator. When $c_{1} =
-\frac{2}{p}$, $p \in {\bf N}$
 eq.(36) corresponds to the para-Fermi oscillator of the $p^{th}$
order\cite{ok}\cite{bon}, and corresponding representations
are finite dimensional. (If we admit a different vacuum , $|0) = |p>$,
there is an infinite tower of  states on it). When $c_{1}>0$, the
representations are  infinite dimensional but different from the para-Bose
oscillator.\\
Notice that the only values of $c_1$ for which eqs.(29) and (36) give
the same oscillator are $c_1 = (0,-2)$;$(2,0)$;$(-1,-1)$,
where the first (second) entry corresponds to the value of $c_1$ in eq.(29)
(eq.(36)),respectively.\\
Simple deformations of the single para-Fermi oscillator \cite{okk}\cite{fv} are
\begin{equation}
\varphi^{(a,b)} (n) = [n]^{(a,b)}_{q} [1 + \frac{1}{2} c_{1}(n-1)]_{q}^{(a,b)}
{}.
\end{equation}
Similarly as in \cite{bmf}, one can write
\begin{equation}
\begin{array}{c}
a \bar{a} - q \bar{a} a = q^{-N} (1 + c_{1} N),\\[4mm]
\varphi (n) = q^{n-1} (S_{0}(n,q^2) + c_{1} S_{1} (n,q^2)).
\end{array}
\end{equation}

Finally, we give a simple interpolation between the para-Bose, the para-Fermi ,
the genon oscillators and all other known deformed single-mode
oscillators\cite{ac}-\cite{jag} :
\begin{eqnarray}
a \bar{a} - q \bar{a} a = p^{-N}  \sum_{k=0}^{\infty} c_{k} N^k ,  \quad p,
q, c_{k} \in {\bf C}, \nonumber \\
\varphi (n) = q^{n-1} \sum_{k=0}^{\infty} c_{k} S_{k}(n,pq) =
\sum_{k=0}^{\infty}
c_{k} \{\sum_{k=0}^{n-1} j^k\}_{(k,pq)}. \label{}
\end{eqnarray}
Finite-dimensional representations are determined with  solutions
 of the equation $\varphi (n_0) = 0$, $n_0 \in {\bf N}$. The commutation
relation, eq.(39), represents the most general single-mode deformed oscillator.
General multimode deformed oscillators are more complicated and some partial
results
are obtained in \cite{fz}.

We conclude with a two remarks.\\{\it Remark 1}. We can define  new operators
$B_{-} = -\frac{1}{2} a^2$ and  $B_{+} = \frac{1}{2} \bar{a}^2$, satisfying the
commutation relations
\begin{equation}
\begin{array}{c}
[ N, B_{\pm} ] = \pm 2 B_{\pm},\\[4mm]
[ B_{+}, B_{-} ] = \frac{1}{4} (\varphi (n+2)\varphi (n+1) - \varphi (n)\varphi
(n-1)),
\end{array}
\end{equation}
and the Casimir operator
\begin{equation}
\begin{array}{c}
C = B_{+} B_{-} + \frac{1}{4} \varphi (n)\varphi (n-1) + C_{0}\\[4mm]
   = B_{-} B_{+} +  \frac{1}{4} \varphi (n+2)\varphi (n+1) + C_{0},\\[4mm]
C_0 = const.
\end{array}
\end{equation}
Using $ \varphi (n)$ from eq.(35) we reproduce $U_{q}(SU(1,1))$ as a
spectrum-generating algebra for the algebra (34). In the limit
$\nu \rightarrow 0$, $\varphi (n)=q^{n-1} S_{0}(n,q^2)$ and the standard
deformation of the $SU_{q}(1,1)$ is reproduced \cite{ck}. The undeformed
-oscillator realization of the $SU(1,1)$ algebra is recovered  with
$\nu \rightarrow 0$, $q \rightarrow 1$, i.e. with $\varphi (n)=n$.\\
{\it Remark 2}. In the same way as in \cite{bmf} we can realize the
supersymmetry algebra $ \{Q_{i}(p),Q_{j}(p)\} = 2 \delta_{ij} H(p)$, $i,j=1,2$
(\cite{ck}\cite{pw}) with the Hermitian operators $Q_{i}(p)$, $p \in {\bf R}$,
given by
\begin{equation}
\begin{array}{c}
Q_1(p) = a^{\dagger} p^{N+1} {\it P_{-}} + a p^{N} {\it P_{+}}\\[4mm]
Q_2(p) = i ( a^{\dagger} p^{N+1} {\it P_{-}} -  a p^{N} {\it P_{+}})\\[4mm]
{\it P_{\pm}} = \frac{1}{2} ( 1 \pm  (-)^N ).
\end{array}
\end{equation}
The hamiltonian $H(p)$ is given as
\begin{equation}
H(p) = |\varphi (N+1)| p^{2(N+1)} {\it P_{-}} + |\varphi (N)| p^{2N} {\it
P_{+}}.
\end{equation}
\bigskip

{\bf Acknowledgments}
This work was supported by the joint Croatian-American contract NSF JF 999 and
the Scientific Fund of the Republic of Croatia.
%
%
%
\bigskip
%

%

\end{document}